\documentclass[12pt,preprint]{aastex}

\slugcomment{Accepted to the ApJL}
\shorttitle{H-D substitution in solid methanol}
\shortauthors{Nagaoka, Watanabe, and Kouchi}

\begin{document}

\title{H-D substitution in interstellar solid methanol: a key route for D enrichment}

\author{Akihiro Nagaoka\altaffilmark{1}, Naoki Watanabe\altaffilmark{1}, 
 and Akira Kouchi\altaffilmark{1}}
\affil{Institute of Low Temperature Science, Hokkaido University, N19-W8, 
Kita-ku, Sapporo, Hokkaido 060-0819, Japan}
\email{watanabe@lowtem.hokudai.ac.jp}

\altaffiltext{1}{Institute of Low Temperature Science, Hokkaido University,
N19-W8, Kita-ku, Sapporo, Hokkaido 060-0819, Japan; 
watanabe@lowtem.hokudai.ac.jp}

\begin{abstract}
Deuterium enrichment of interstellar methanol is reproduced experimentally 
for the first time via grain-surface H-D substitution in solid methanol at 
an atomic D/H ratio of 0.1. Although previous gas-grain models successfully 
reproduce the deuterium enrichments observed in interstellar methanol 
molecules (D/H of up to 0.4, compared to the cosmic ratio of $\sim 
$10$^{-5})$, the models exclusively focus on deuterium fractionation 
resulting from the successive addition of atomic hydrogen/deuterium on CO. 
The mechanism proposed here represents a key route for deuterium enrichment 
that reproduces the high observed abundances of deuterated methanol, 
including multiple deuterations. 
\end{abstract}


%\newpage 

\section{Introduction}

Recent astronomical observations (Turner 1990; van Dishoeck et al. 1995; Loinard et al. 2000; Parise et al. 2002; Parise et al. 2004; Crovisier et al. 2004) have revealed that the abundances of 
some deuterated interstellar molecules are up to 4 orders of magnitude 
higher than the cosmic D/H ratio of 1.5~$\times $~10$^{-5}$ \citep{Lin98}. In the case 
of methanol molecules in particular, even doubly and triply deuterated 
isotopologues have been detected in the vicinity of low-mass protostars (Parise et al. 2002; 2004). 
Deuterium fractionation in interstellar molecules has long been a topic of 
interest in interstellar chemistry, and the fractionation processes in 
molecular clouds have attracted attention from both astronomers and earth 
and planetary scientists as a potential explanation for the high D 
fractionations observed in our solar system (for a review, see Robert, Gautier,
{\&} Dubrulle 2000). Although many theoretical 
models of interstellar D fractionation have been proposed, pure gas-phase 
models have difficulties reproducing the high observed D enrichments, 
particularly multiple deuterations such as D$_{2}$CO and CD$_{3}$OH. It has 
thus been recognized that grain-surface chemistry is a necessary process for 
the production of multiply deuterated species. This is a readily 
understandable result for formaldehyde and methanol, which are likely to be 
produced efficiently on grains by tunneling reactions under the conditions 
of molecular clouds (Watanabe, Shiraki, {\&} Kouchi 2003; Watanabe et al. 2004; Hidaka et al. 2004). Some of the proposed gas-grain models therefore 
provide reasonable estimates of the high D enrichment observed in 
interstellar formaldehyde and methanol (Tielens 1983; Charnley, Tielens, {\&} Rodgers 1997; Caselli et al. 2002; Aikawa, Ohashi, {\&} Herbst 2003).
 Recently, a model using direct 
master equations has also been successful in reproducing the D enrichment of 
methanol, including triple deuteration, on the assumption of a high atomic 
D/H ratio of $\sim $0.3 in the accreting gas \citep{Sta03}. In such models, the high 
atomic D/H ratio is produced by ion--molecule reactions in the gas phase 
followed by hydrogenation and deuteration of CO on grain surfaces, leading 
to D enrichment of methanol. For instance, the atomic D/H ratio of up to 0.1 
was achieved in the gas-phase model by Roberts, Herbst {\&} Millar (2002). 
However, the results of these models are 
somewhat ambiguous due to the requirement of the very high atomic D/H ratio 
(0.3--1) and the lack of information regarding grain-surface reactions, 
including information such as reaction channels, activation energies, and 
the rates of hydrogenation (deuteration). It is therefore highly desirable 
to clarify experimentally the role of surface reactions in D fractionation. 

Here we report the experimental results on the exposure of solid CO to cold H and D atoms.  The observed D fractionation in methanol is experimentally 
reproduced by grain-surface reactions at an atomic D/H ratio of 0.1, revealing 
that the 
key reaction for D fractionation is the H-D substitution in solid CH$_{3}$OH 
with atomic D rather than successive addition of H and D on CO. As the 
interaction of D atoms with methanol has not been treated in previous 
models, this may offer an explanation for the observed abundance of 
deuterated methanol isotopologues in molecular clouds.

\section{Experimental}

Experiments were performed using the apparatus for surface reactions in 
astrophysics (ASURA) system described previously \citep{Wat04}. The sample solid was 
produced by vapor deposition on an aluminum substrate at 10~K in an 
ultra high vacuum chamber (typically 1~$\times $~10$^{-10}$~torr). Atomic H 
and D were produced by a microwave-induced plasma in a Pyrex tube and 
transferred via a series of polytetrafluorethylene (PTFE) tubes to the 
target. A deflector was mounted at the end of the PTFE tube to eliminate 
charged particles and metastable atoms that may escape from the plasma. 
Atomic beam were cooled in the PTFE tube, which was tightly sheathed with a 
copper tube connected to an He refrigerator. The kinetic temperature of 
atoms was set at 30~K. For the simultaneous exposure of solid CO to H and D atoms, H$_{2}$ and D$_{2}$ gases were mixed in a stainless steel bottle at 
a certain ratio and introduced into the Pyrex tube. The flux of atoms was 
measured using a quadrupole mass spectrometer installed close to the 
substrate and was maintained constant through the measurements. The infrared 
absorption spectra of the sample solid were measured during exposure to 
atoms by Fourier transform infrared spectroscopy (FTIR) with a resolution of 
4~cm$^{-1}$. 

\section{Results and Discussion}

A pure solid CO sample of approximately 10 monolayers was exposed to cold H 
and D atoms simultaneously at a D/H ratio of 0.1. This atomic ratio was 
adopted as a representative value that can be achieved in the gas-phase model 
\citep{Rob02}. Fig. 1 shows the infrared 
absorption spectrum of the initial solid and the variation in the spectrum 
upon exposure to H and D atoms. Formaldehyde (H$_{2}$CO), methanol 
(CH$_{3}$OH), HDCO, D$_{2}$CO, and Me-$d_{n}$-OH clearly appear, with the 
consumption of CO. Here, Me-$d_{n}$- represents CH$_{2}$D-, CHD$_{2}$-, and 
CD$_{3}$- for n = 1-3, respectively. It should be noted that CH$_{3}$OD and  
Me-$d_{n}$-OD were not detected. The band for OD stretch of CH$_{3}$OD locates 
at around 2430~cm$^{-1}$ with the band strength being 1.4 times larger 
than that of the CO-stretch of CH$_{3}$OH at 1043~cm$^{-1}$. We did not observe 
the peak there. Taking account of the detection limit of the FTIR, the 
CH$_{3}$OD yield should be 1 order of magnitude less than those of Me-$d_{n}
$-OH at the maximum. This is 
consistent with the much lower abundances of Me-OD compared to 
Me-$d_{n}$-OH in molecular clouds \citep{Par04}. Using pure solid samples of 
deuterated methanol, the band strengths relative to that 
of pure CH$_{3}$OH at 1043~cm$^{-1}$ and 10~K were determined to be 0.10, 
0.18, 0.47, and 0.35 for CH$_{2}$DOH (1330~cm$^{-1})$, CHD$_{2}$OH 
(950~cm$^{-1})$, CD$_{3}$OH (988~cm$^{-1})$, and CD$_{3}$OD (982~cm$^{-1})$, 
respectively. The variation in column densities for CO, CH$_{3}$OH and 
Me-$d_{n}$-OH and the sum of deuterated methanol are plotted in Fig. 2. 
Although H$_{2}$CO, D$_{2}$CO and HDCO were also observed, these are not 
plotted because of the unknown band strengths of HDCO and D$_{2}$CO 
molecules. Assuming an H number density of 1~cm$^{-3}$ and an accreting atomic 
D/H ratio of 0.1, the D (H) atom fluences in a 10~K molecular cloud will be 
1.3~$\times $~10$^{16}$ (1.3~$\times $~10$^{17})$, 6.5~$\times $~10$^{16}$ 
(6.5~$\times $~10$^{17})$, and 1.3~$\times $~10$^{17}$~(1.3~$\times 
$~10$^{18})$~cm$^{-2}$ over 10$^{5}$, 5~$\times $~10$^{15}$ and 10$^{6}$~yr, 
respectively. 
In the present experiments, these fluences approximately correspond to 
exposure times of 10, 35, and 70~minutes, respectively. The ratios of deuterated 
methanol to the amount of remaining CH$_{3}$OH, denoted in the form 
(CH$_{2}$DOH/CH$_{3}$OH, CHD$_{2}$OH/CH$_{3}$OH, CD$_{3}$OH/CH$_{3}$OH), 
after exposure times of 10, 35, and 70~min are (0.16, 0, 0), (0.09, 0.05, 
0.01) and (0.11, 0.13, 0.03), respectively. These values are closely 
consistent with the observations of (0.3$\pm$0.05, 0.06$\pm$0.01, 0.014$\pm$0.006) (Parise et al. 2004; Aikawa et al. 2005) and clearly show that the D 
fractionation in methanol can be achieved via grain-surface reactions at the 
accreting atomic D/H ratio of 0.1. It should also be noted that this 
estimate is quite general and depends heavily on the conditions in the 
molecular cloud. Nevertheless, grain-surface D fractionation of CH$_{3}$OH 
in molecular clouds has been demonstrated to proceed more efficiently than 
predicted by previous models. 

There are two possible routes to produce deuterated methanol in the present 
experiment: successive hydrogenation/deuteration of CO and the H-D 
substitution in CH$_{3}$OH. The profile of the variation curves of 
Me-$d_{n}$-OH in Fig.2 (a) is apparently dominated by the latter process 
rather than the former one. As the exposure time increases, the deuterated 
isotopologues appear in the following sequence, consuming the 
lower deuterated methanol from $d_{0}$ to $d_{3}$: CH$_{3}$OH~$\to 
$~CH$_{2}$DOH~$\to $~CHD$_{2}$OH~$\to $~CD$_{3}$OH. Additionally, the curves 
of the CH$_{3}$OH and CH$_{2}$DOH yields have maxima. These features clearly 
differ from those expected for successive addition of atoms, indicating the 
H-D substitution in CH$_{3}$OH. If the deuterated isotopologues are mainly 
produced by successive hydrogenation/deuteration and no exchange occurs in 
methanol, the curves of CH$_{3}$OH and Me-$d_{n}$-OH should not have 
maxima. Furthermore, the ratios of Me-$d_{2,3}$-OH/CH$_{2}$DOH obtained are 
much larger than those in models, including only the former process, with the 
accreting D/H ratio of 0.1 (Stantcheva {\&} Herbst 2003; Caselli et al. 2002).

The process of H--D substitution in CH$_{3}$OH was further examined by 
exposing pure solid methanol (CH$_{3}$OH) to a single D atom beam and to an 
H + D atomic beam (D/H ratio of 1) at 10~K. Fig. 3 shows the infrared 
absorptio spectrum of the initial pure solid CH$_{3}$OH and the variation in 
the spectrum upon exposure to D atoms. When D$_{2}$ and HD molecules were 
deposited on the sample, no changes in the spectrum were found, indicating 
that CH$_{3}$OH is consumed upon exposure to cold D atoms. Fig. 4 shows the 
variation in column densities of CH$_{3}$OH and deuterated isotopologues for 
the D-exposure as a function of exposure time. For the exposure to H + D 
atoms, only the decrease in CH$_{3}$OH is plotted, because the variations of 
isotopologues are very similar to those for the single D-exposure. The 
deuterated isotopologues are clearly formed with exposure time in the 
following sequence: CH$_{3}$OH~$\to $~CH$_{2}$DOH~$\to $~CHD$_{2}$OH~$\to 
$~CD$_{3}$OH. The production of Me-OD (OD stretching) was not observed 
again. These features are consistent with the measurements described above. 
The deuteration of CD$_{3}$OH was also investigated, but again no change in 
the spectrum was observed. This may be due in part to the larger 
dissociation energy of CH$_{3}$OH to CH$_{3}$O~+~H than that of CH$_{3}$OH 
to CH$_{2}$OH~+~H \citep{Bau91}, as the O-H bond is more stable than the C-H bond. This is the first clear evidence that solid CH$_{3}$OH is deuterated 
through CH$_{2}$DOH and CHD$_{2}$OH to CD$_{3}$OH by exposure to D atoms at 
10~K. Most of the grain models focus on the successive hydrogenation/
deuteration of CO to produce the deuterium fractionation in methanol, 
and they partly succeed in reproducing the observations. However, the present 
results suggest that the H-D substitution is more efficient than the 
hydrogenation/deuteration process and thus should be included in the reaction 
network. In the previous models, the deuteration rates may have been 
overestimated because of the use of a simple potential for the tunneling 
reactions.  

Exposure of solid CH$_{3}$OD, CD$_{3}$OH, CH$_{2}$DOH, CHD$_{2}$OH, and 
CD$_{3}$OD to H atoms at 10~K similarly did not induce any change in the 
infrared absorption spectrum, suggesting that deuterated methanol is very 
stable and that the hydrogenation of deuterated methanol is inhibited. 
Therefore, the present data show that the exposure of CH$_{3}$OH to D atoms 
represents a previously unknown mechanism for the efficient formation of 
deuterated methanol. 

As shown in Fig. 3(b), exposure to D atoms results in the formation of 
methyl-deuterated methanol isotopologues, CH$_{2}$DOH, CHD$_{2}$OH, and 
CD$_{3}$OH, but no other isotopologues (e.g., CH$_{3}$OD, CH$_{2}$DOD, and 
CHD$_{2}$OD). Therefore, H--D substitution proceeds by either or both of the 
following reactions: 

CH$_{3}$OH~$\to $~CH$_{2}$OH~$\to $~CH$_{2}$DOH~$\to $~CHDOH~$\to 
$~CHD$_{2}$OH~$\to $~CD$_{2}$OH~$\to $~CD$_{3}$OH (1)

CH$_{3}$OH~$\to $~CH$_{2}$DOH~$\to $~CHD$_{2}$OH~$\to $~CD$_{3}$OH (2)

Reaction (1) is a repetition of the formation of a hydroxymethyl radical by 
H abstraction from methanol and subsequent D addition to the hydroxymethyl 
radical to form deuterated methanol, e.g., CH$_{3}$OH~+~D~$\to 
$~CH$_{2}$OH~+~HD and CH$_{2}$OH~+~D~$\to $~CH$_{2}$DOH. For formaldehyde,
the importance of this process was predicted by \citet{Tie83}. Reaction (2) 
represents direct H--D exchange, such as CH$_{3}$OH~+~D~$\to 
$~CH$_{2}$DOH~+~H. Although it is unclear which process is dominant, 
deuteration rate depends mainly on the D atoms' flux because the rate of 
deuteration for the single D beam is very similar to that for the H + D beam 
at the D/H ratio of 1, as shown in Fig. 4. The nondetection of Me-OD in 
the present experiments can be explained by the H-abstraction/D-
addition (scheme of reaction (1)). The H-
abstraction on the methyl-side by either H or D has a lower barrier than that in -OH 
\citep{Ker04}.

In conclusion, D enrichment in methanol at levels comparable to observations 
was demonstrated to be possible through surface reactions between cold D 
atoms and methanol at 10~K with an accreting atomic D/H ratio of 0.1. Since we 
have not tried another D/H ratio, it is not clear whether or not the present ratio of 
0.1 is the minimum value for reproducing the observed enrichments. However, the 
fractionation levels for the different ratios can be roughly estimated from 
Fig. 2 (b) if we neglect the effect of hydrogen atoms. That is, the fluence of 
D atoms for the ratio of 0.05 corresponds to half of that for 0.1 at a certain 
exposure time. In this estimation, the ratio of 0.05 also seems to reproduce 
the observations fairly well. The deuteration proceeds mainly via the H--D 
substitution in solid methanol rather than successive hydrogenation and 
deuteration of CO. Although the fractionation level must decrease with the 
decrease of the atomic D/H ratio, the contribution of H--D substitution to 
D-enrichment would become more significant than successive 
hydrogenation/deuteration of CO because initially more CH$_{3}$OH is rapidly 
produced. On the other hand, at higher accreting D/H ratios, 
successive hydrogenation/deuteration of CO will be enhanced because the 
chance of D addition on CO relatively increases. Unfortunately, in the present 
work, it is difficult to compare the substitution with the H and D additions 
more quantitatively. 

Other deuterated molecules, such as ammonia, formaldehyde, and water, are 
also present in molecular clouds and/or comets, and the degree of 
deuteration among the different molecules has been observed to differ 
substantially. Although we investigated NH$_{3}$~+~D~ system, no reactions occurred in the exposure of solid 
NH$_{3}$ to cold D atoms at below 15~K. H. Hidaka, N. Watanabe, {\&} A. Kouchi, (2005, in preparation) also found 
experimentally that no deuteration of H$_{2}$O occurs under D exposure at 
10--20~K, even under fluences of up to 5~$\times $~10$^{18}$~cm$^{-2}$. 
This is consistent with the findings of Parise et al. (2003), who did
not detect HDO along the line of sight of low-mass protostars.
However, the deuteration of formaldehyde is likely to proceed under D 
exposure, and significant amounts of HDCO and D$_{2}$CO were observed in the 
present experiments. Further experiments on formaldehyde should therefore 
prove helpful for developing a comprehensive understanding of the 
deuteration mechanism of interstellar molecules.

\clearpage

\begin{figure}
\epsscale{.50}
\plotone{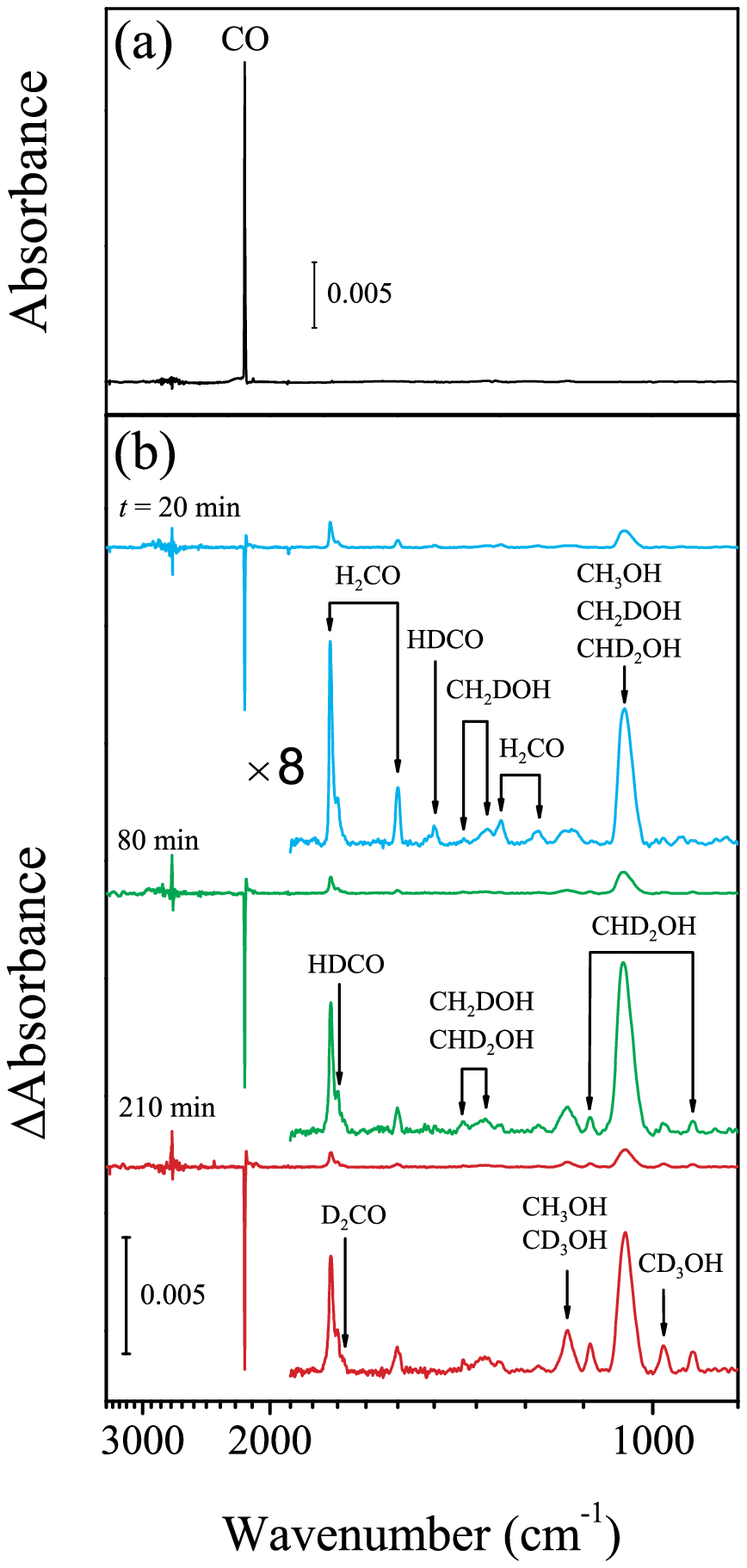}
\caption{(a) Infrared absorption spectrum of initial solid CO 
deposited at 10~K. (b) Spectral change after simultaneous exposure to H and 
D atoms for 20, 80, and 210~min. Spectra were obtained by subtracting the 
initial (non-exposure) absorption spectrum from the irradiated spectra. 
Peaks below and above the baseline represent decreases and increases in 
absorbance, respectively. The newly formed molecules are denoted by arrows. 
Spikes at around 2685~cm$^{-1}$ are noise caused by vibration of the He 
refrigerator. The band above the baseline at 1040~cm$^{-1}$ represents the 
superposition of CH$_{3}$OH, CH$_{2}$DOH, and CHD$_{2}$OH in the CO 
stretching mode. \label{fig1}}
\end{figure}
\clearpage

\begin{figure}
\epsscale{.80}
\plotone{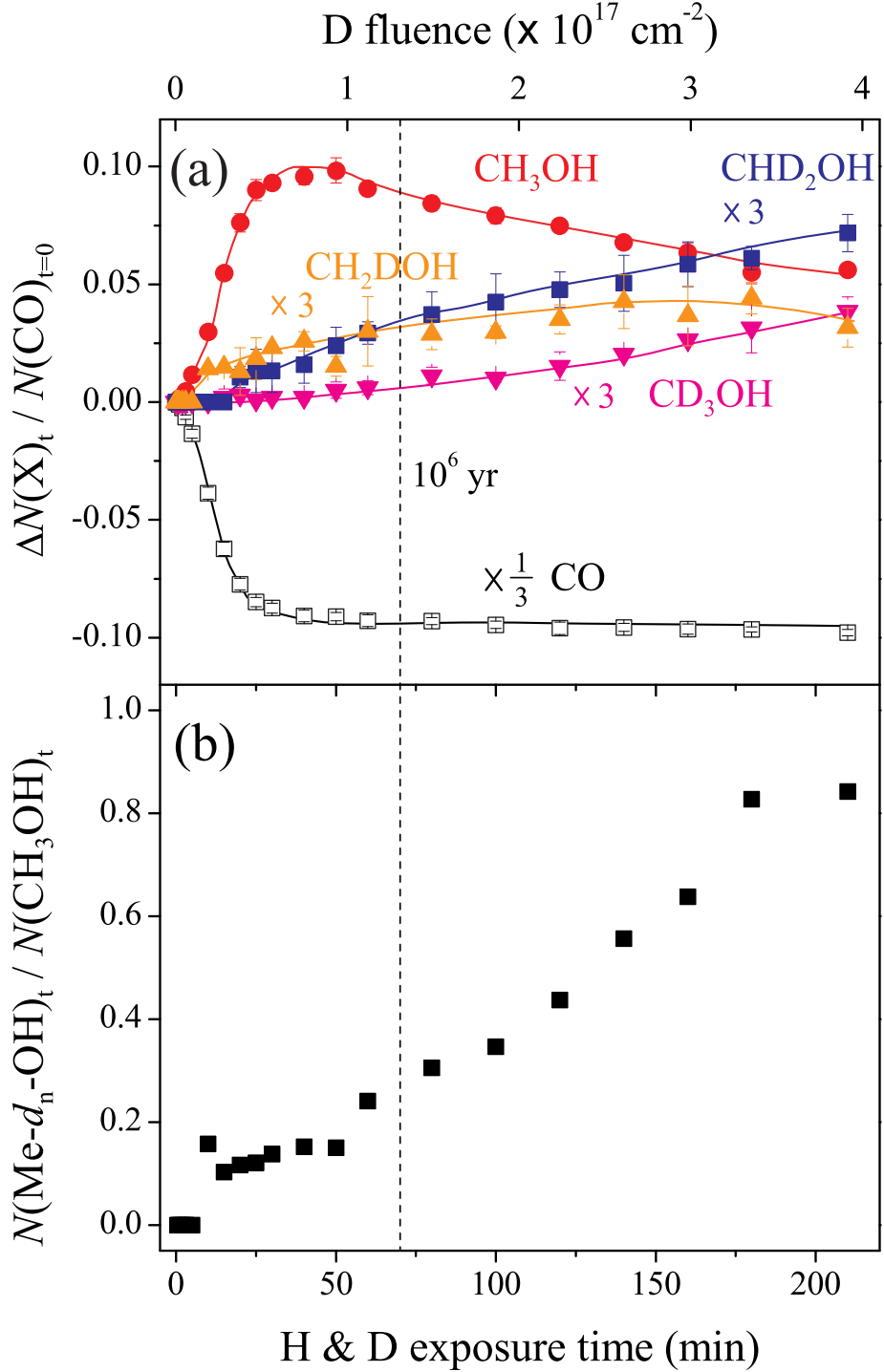}
\caption{(a) Variation in column densities for CO, CH$_{3}$OH, 
CH$_{2}$DOH, CHD$_{2}$OH, and CD$_{3}$OH with H--D exposure time. Column 
densities are normalized to that of initial CO. Intensities of error bars 
represent statistical error. The CH$_{3}$OH yield was derived from the band 
area at 1040~cm$^{-1}$ after subtracting the increase in CH$_{2}$DOH and 
CHD$_{2}$OH in the same band. Column densities for CH$_{2}$DOH, CHD$_{2}$OH, 
and CD$_{3}$OH are magnified to 3 times, and column density for CO is divided by 
3. Solid lines are guides. (b) Ratio of column density for the sum of 
deuterated methanol to that for the remaining CH$_{3}$OH.\label{fig2}}
\end{figure}
\clearpage
\begin{figure}
\epsscale{.80}
\plotone{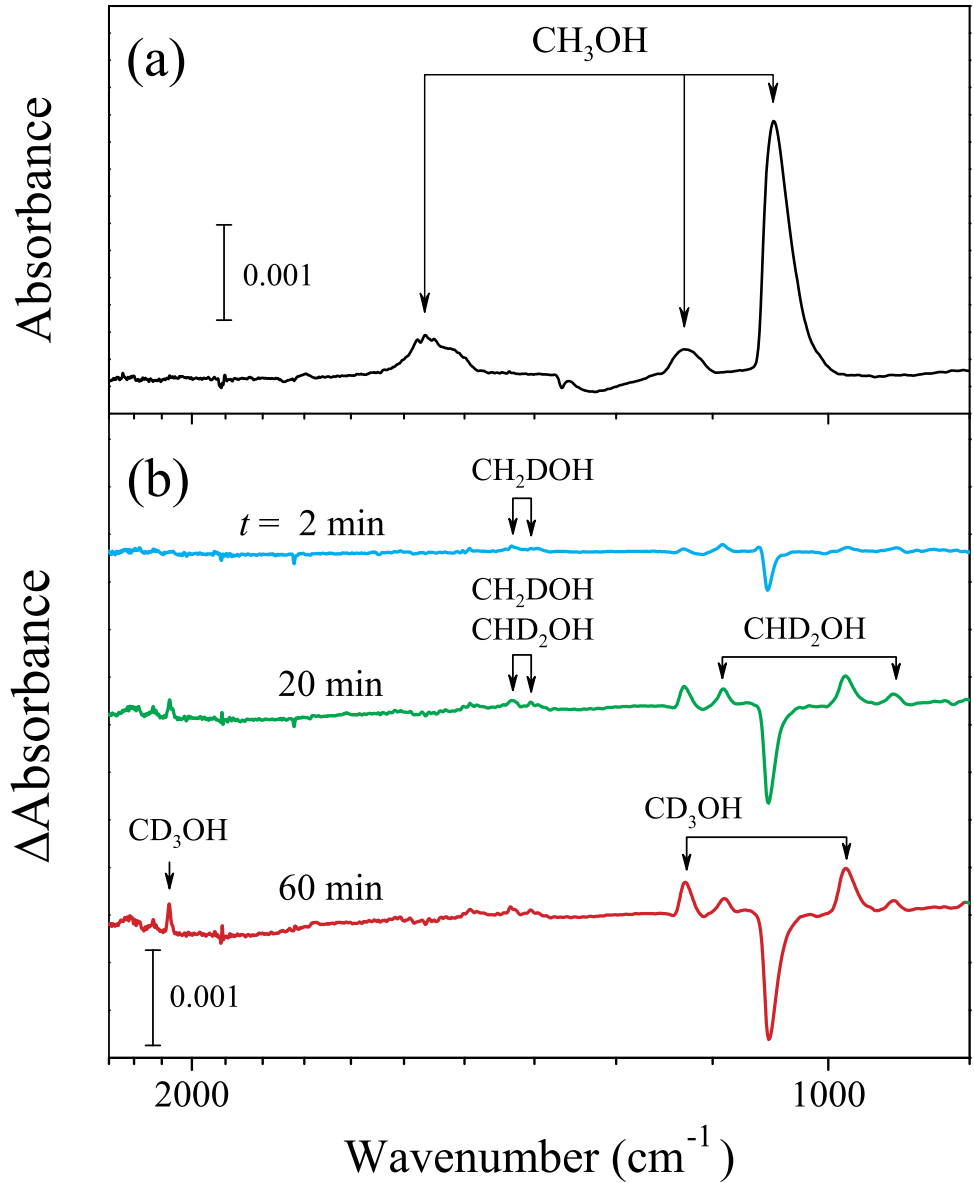}
\caption{(a) Infrared absorption spectrum of initial solid 
CH$_{3}$OH deposited at 10~K. (b) Spectral change after D exposure for 2, 
20, 60~min. Spectra were obtained by subtracting the initial (non-exposure) 
absorption spectrum from the D-exposed spectra. Peaks below and above the 
baseline represent decreases and increases in absorbance, respectively. 
Spikes at 1905 and 1723~cm$^{-1}$ are noise caused by vibration of the He 
refrigerator. The band below the baseline at 1049~cm$^{-1}$ represents the 
superposition of a decrease (CH$_{3}$OH) and increase (CH$_{2}$DOH and 
CHD$_{2}$OH) in the CO stretching mode. \label{fig3}}
\end{figure}
\clearpage
\begin{figure}
\epsscale{.80}
\plotone{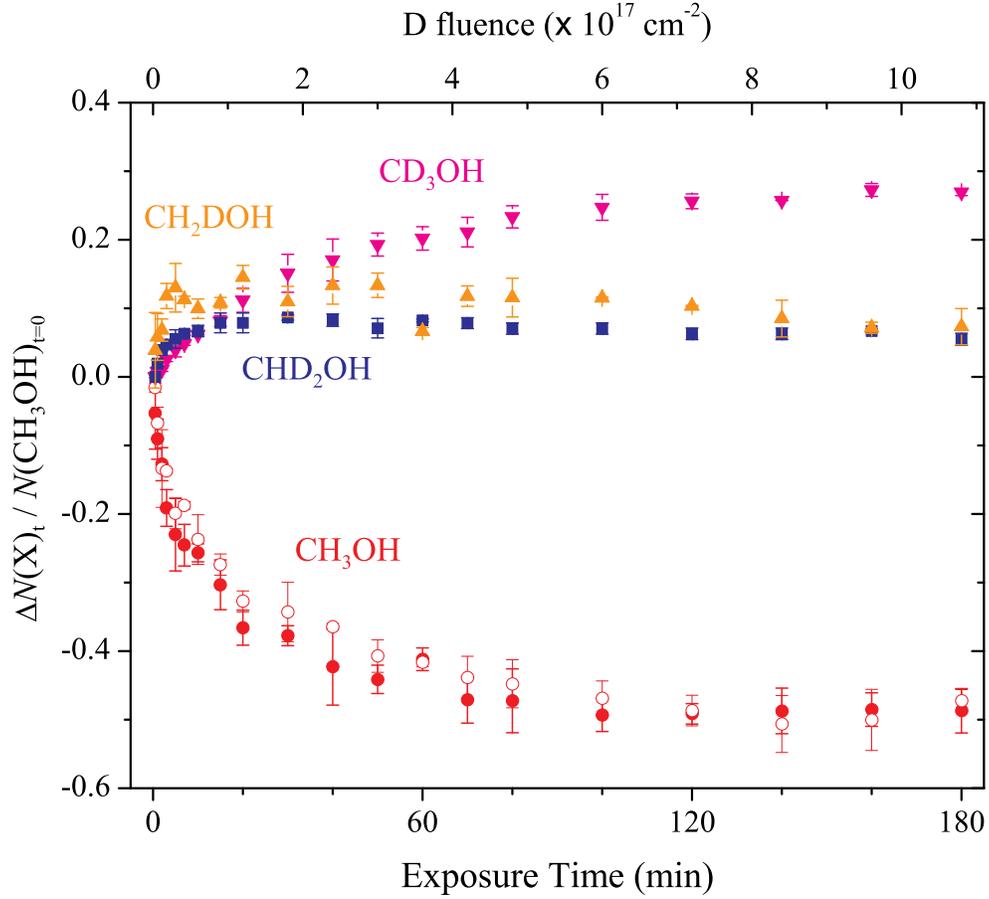}
\caption{Variation in column densities for CH$_{3}$OH and 
isotopologues normalized to initial CH$_{3}$OH with D-exposure. For exposure 
to H + D atoms, only the decrease of CH$_{3}$OH is plotted: the red open circles 
represent H and D fluxes of 1~$\times $~10$^{14}$~cm$^{-2~}$s$^{-1}$. The 
decrease in CH$_{3}$OH was derived from the band area at 1049~cm$^{-1}$ 
after subtracting the increase in CH$_{2}$DOH and CHD$_{2}$OH in the same 
band.\label{fig4}}
\end{figure}
\clearpage

\end{document}